\documentclass[prl,twocolumn,superscriptaddress]{revtex4-2}
\usepackage{amssymb}
\usepackage{amsfonts}
\usepackage{bbm}
\usepackage{mathrsfs}
\usepackage{graphics,graphicx,epsfig,bm,amsmath,amsthm,amssymb}
\usepackage{bm}
\usepackage{bbm}
\usepackage{longtable}
\usepackage{multirow}
\usepackage{array}
\usepackage{color}
\usepackage{cases}
\usepackage{booktabs }
\usepackage[usenames,dvipsnames]{xcolor}
\newcommand{\ket}[1]{\left| #1\right\rangle}
\usepackage{xcolor}
\usepackage[none]{hyphenat}
\usepackage{float}
\usepackage{amsmath}

\usepackage{verbatim}
\usepackage{makecell}
\usepackage{amsmath,bm}
\usepackage{physics}
\usepackage{simplewick}
\usepackage{slashed}
\usepackage[a4paper,colorlinks=true,
linkcolor=blue,citecolor=blue,
pdfauthor={ },
pdftitle={ },
pdfsubject={ },
pdfkeywords={ }]{hyperref}

\begin{document}

\title{Exotic-interaction searches on small scales via near-threshold enhancement}

\author{Runqi Kang}
\affiliation{Laboratory of Spin Magnetic Resonance, School of Physical Sciences, Anhui Province Key Laboratory of Scientific Instrument Development and Application, University of Science and Technology of China, Hefei 230026, China}
\affiliation{Hefei National Laboratory, University of Science and Technology of China, Hefei 230088, China}

\author{Xing Rong}
\email{xrong@ustc.edu.cn}
\affiliation{Laboratory of Spin Magnetic Resonance, School of Physical Sciences, Anhui Province Key Laboratory of Scientific Instrument Development and Application, University of Science and Technology of China, Hefei 230026, China}
\affiliation{Hefei National Laboratory, University of Science and Technology of China, Hefei 230088, China}
\affiliation{Hefei National Research Center for Physical Sciences at the Microscale, Hefei 230026, China}
\affiliation{State Key Laboratory of Ocean Sensing and School of Physics, Zhejiang University, Hangzhou 310058, China}
\affiliation{Institute of Quantum Sensing, Institute of Fundamental and Transdisciplinary Research and Zhejiang Key Laboratory of R\&D and Application of Cutting-edge Scientific Instruments, Zhejiang University, Hangzhou 310058, China}

\begin{abstract}

Exotic interactions between fermions mediated by new bosons beyond the Standard Model may hold the key to several fundamental conundrums on the frontier of physics.
However, laboratory searching for exotic interactions on small length scales is fundamentally held back by the short force range.
Here we propose that the force range of exotic interactions tends to infinity
  when the oscillation frequency of exotic interactions approaches the mass of the new bosons,
  i.e., the new bosons become nearly on-shell.
This is named as the near-threshold enhancement.
Through the near-threshold enhancement, even fermions at distances larger than the original force range can make considerable contributions to exotic interactions.
Therefore, the size of the experimental apparatus can break the limitation of the force range, 
  and thus both the signal of exotic interactions and the sensitivity of sensors can be greatly enhanced.
We also propose a method to search for the exotic interactions in the mass range between 80 $\mu$eV and 800 $\mu$eV taking advantage of the near-threshold enhancement.
For the coupling $g_{\rm A}^{\rm e}g_{\rm A}^{\rm e}$, 
  we expect an improvement on its upper bounds of 10 orders of magnitude at 800 $\mu$eV.
This method can be further extended to enhance the search for other types of exotic interactions and boost the study of new physics beyond the Standard Model.

\end{abstract}

\maketitle

The existence of new bosons beyond the Standard Model are indicated by numerous anomalous phenomena,
  such as the velocity discrepancy of galaxies \cite{1906Poincare,1933Zwicky}, 
  the cosmic dipole problem \cite{2023Han}, 
  and the puzzling deviation between the theoretical prediction and measurement of the many properties of fundamental particles \cite{2021Abi,2021Cazzaniga,2022Aaltonen,2022Thomas}.
Exotic interactions are predicted to rise between ordinary fermions from the exchange of these new bosons \cite{1984Moody}.
With two spins and two momenta, there are 16 independent scalars including all possible spin configurations,
  $\mathcal{O}_{i} = \mathcal{O}_{i}(\bm{\sigma}_{1},\bm{\sigma}_{2},\bm{q},\bm{P})(i=1,2,\dots 16)$,
  where $\bm{\sigma}_{1,2}$ are the spins of the fermions,
  $\bm{P}$ is the average momentum of a fermion
  and $\bm{q}$ is the momentum transmission during the scattering.
These operators give rise to 16 exotic interactions,
  $V_{\rm i} = f_{\rm i}\mathcal{V}_{\rm i}e^{-r/\lambda}(i=1,2,\dots 16)$,
  where $f_{\rm i}$ is the strength of the interactions,
  $\mathcal{V}_{\rm i}$ refers to the polynomial part of the interactions,
  $r$ is the distance between the fermions,
  $\lambda = m_{\rm X}^{-1}$ is the force range,
  and $m_{\rm X}$ is the mass of the new bosons \cite{2006Dobrescu} 
  (see more details in Sec. \uppercase\expandafter{\romannumeral1} of Supplemental Material \cite{spp}).
The force range $\lambda$ is also the Compton wavelength of a new boson at rest.
The exotic interactions were later sorted into 9 types according to the couplings between fermions and the new bosons \cite{2019Fadeev,2025Cong}.
For example, the pseudovector/pseudovector interaction induced by $Z'$ bosons is
\begin{equation}
  V_{\rm AA} = -g_{\rm A}^{1}g_{\rm A}^{2}(\mathcal{V}_{2}+\frac{M^{2}}{m_{\rm X}^{2}}\mathcal{V}_{3})e^{-r/\lambda},
\end{equation}
  where $M = m_{\psi_{1}}m_{\psi_{2}}/(m_{\psi_{1}}+m_{\psi_{2}})$ is the reduced mass of the two-fermion system,
  $m_{\psi_{1,2}}$ are the mass of fermion 1 and 2, respectively,
  and $g_{\rm A}^{1,2}$ is the pseudovector coupling between the fermions and the new bosons
  (see more details in Sec. \uppercase\expandafter{\romannumeral2} of Supplemental Material \cite{spp}).

Lots of efforts have been devoted to search for exotic interactions.
Since most of the exotic interactions can be regarded as pseudomagnetic fields, 
  a widely applied strategy is to explore them using high-precision magnetometers, 
  such as superconducting quantum interference devices \cite{2013Tullney}, 
  spin-exchange-relaxation-free (SERF) magnetometers \cite{2019Kim}, and nitrogen-vacancy (NV) centers \cite{2018Rong,2021Jiao}.
Force detectors like torsion pendulums have also been used to search for exotic interactions \cite{2011Hoedl,2015Terrano}.
Despite the plentiful efforts, 
  a fundamental force range limitation is always hindering the searches for exotic interactions on small length scales.
Since the exotic interactions decay exponentially with the distance between the fermions,
  the size of the experimental setups is limited to the order of $\lambda$.
As a result, both the strength of the signal and the sensitivity of the sensor degrade with $\lambda$ rapidly.
It is noteworthy that a similar problem also hinders the searches for other kinds of interactions, 
  such as forces mediated by quadratically coupled new particles or neutrinos.
For these forces, some theoretical studies proposed that they can be enhanced if the fermions were in a background of the mediators \cite{2024Tilburg,2023Ghosh}.

In this work, we propose that exotic interactions can be brought to a near-threshold regime as the frequency of their oscillation approaches the mass of the new bosons.
In such a regime, the new bosons become nearly on-shell and the exponential decay of exotic interactions vanishes.
By taking advantage of such an effect, 
  the strength of exotic interactions can remain considerable even when the distance between fermions is larger than $\lambda$,
  and the size of the experimental setups can break the limitation of $\lambda$.
As a result, both the signal of the exotic interactions and the sensitivity of the sensor can be greatly enhanced.
We further propose an experimental setup for searching for exotic interactions.
For the coupling $g_{\rm A}^{\rm e}g_{\rm A}^{\rm e}$, 
  the expected constraints at $m_{\rm X} = 800\ \mu$eV exceed the previous results by around 10 orders of magnitude.


Herein we take the pseudovector/pseudovector interaction induced by $Z'$ bosons for example to derive the near-threshold enhancement.
In the quantum field theory (QFT) the interaction Hamiltonian contributed by the pseudovector coupling between $Z'$ bosons and fermions is
\begin{equation}
  \mathcal{H}_{\rm I}^{Z'} = g_{\rm A}^{\psi}Z'_{\mu}\bar{\psi}\gamma^{\mu}\gamma^{5}\psi,
  \label{equ2}
\end{equation}
  where $Z'_{\mu}$ and $\psi$ denotes the $Z'$ bosonic field and the fermionic field, respectively \cite{1984Moody,2019Fadeev,2025Cong}.
\begin{figure}
  \centering
  \includegraphics[width=1\columnwidth]{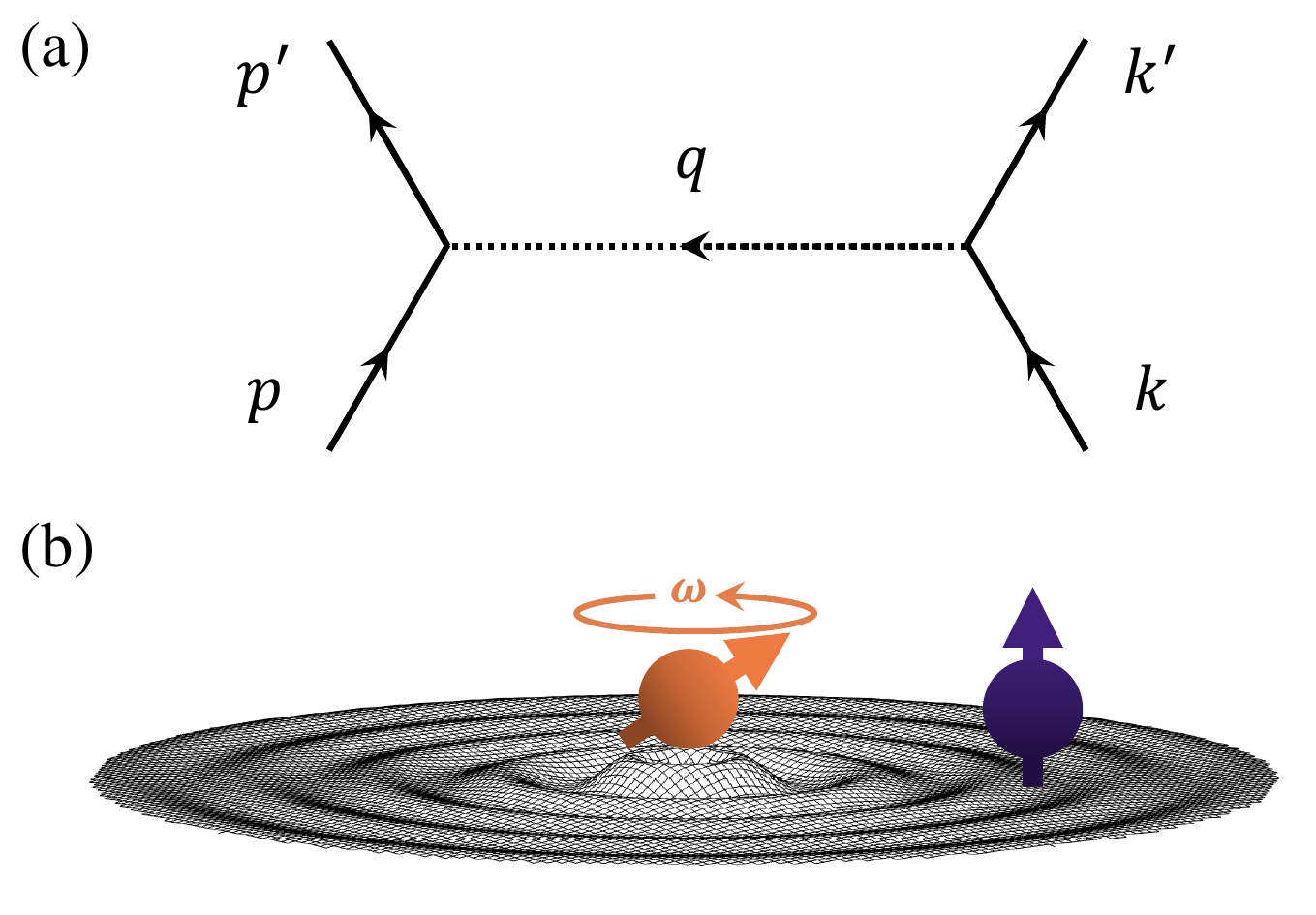}
  \caption{ 
      (a) Feynman diagram of the hypothesized exotic interactions. 
      (b) Schematic diagram of the time-dependent exotic interaction.
          Two fermions in the standard model (orange and purple) interact with each other through exchanging new bosons beyond the standard model.
          Since the new bosonic field is driven periodically by the precessing fermion, a new bosonic wave emerges and spreads out (black).
  }
  \label{Figure1}
\end{figure}
Figure \ref{Figure1}(a) shows a specific case, the two-fermion scattering reaction, 
  $\textrm{fermion}_{1}(p)+\textrm{fermion}_{2}(k) \rightarrow \textrm{fermion}_{1}(p')+\textrm{fermion}_{2}(k')$,
  where $p,k$ are the 4-momentum vectors of the incident fermions while $p',k'$ are the 4-momentum vectors of the scattered fermions.
The leading contribution comes from the $\mathcal{H}_{\rm I}^{2}$ term of the $S$-matrix:
\begin{equation}
  \begin{aligned}
    \langle iT_{\rm AA}\rangle  &=\\
      \bra{\bm{p}'\bm{k}'}&\mathcal{T}[\int d^{4}x(-i\mathcal{H}_{\rm I}(x)) \int d^{4}y(-i\mathcal{H}_{\rm I}(y))]\ket{\bm{pk}},
  \end{aligned}
  \label{Eq3}
\end{equation}  
  where $\bm{p},\bm{k},\bm{p}',\bm{k}'$ are the 3-momenta, and $\mathcal{T}$ is the time ordering operator.
In the nonrelativistic limit, the QFT calculation gives
\begin{equation}
  \begin{aligned}
    \langle iT_{\rm AA}\rangle = -ig_{\rm A}^{\psi_{1}}g_{\rm A}^{\psi_{2}}4m_{\psi_{1}}m_{\psi_{2}}
                                  \frac{\mathcal{O}_{2}+\frac{M^{2}}{m_{\rm X}^{2}}\mathcal{O}_{3}}{q^{2}-m_{\rm X}^{2}}\\
                                  \times(2\pi )^{4}\delta^{4}(p'+k'-p-k),
    \label{equ4}
  \end{aligned}
\end{equation} 
  where $\mathcal{O}_{2} = \bm{\sigma}_{1}\cdot\bm{\sigma}_{2}$, $\mathcal{O}_{3} = (\bm{\sigma}_{1}\cdot q)(\bm{\sigma}_{2}\cdot q)/M^{2}$, and $q = p'-p$.

\begin{figure}[b]
  \centering
  \includegraphics[width=1\columnwidth]{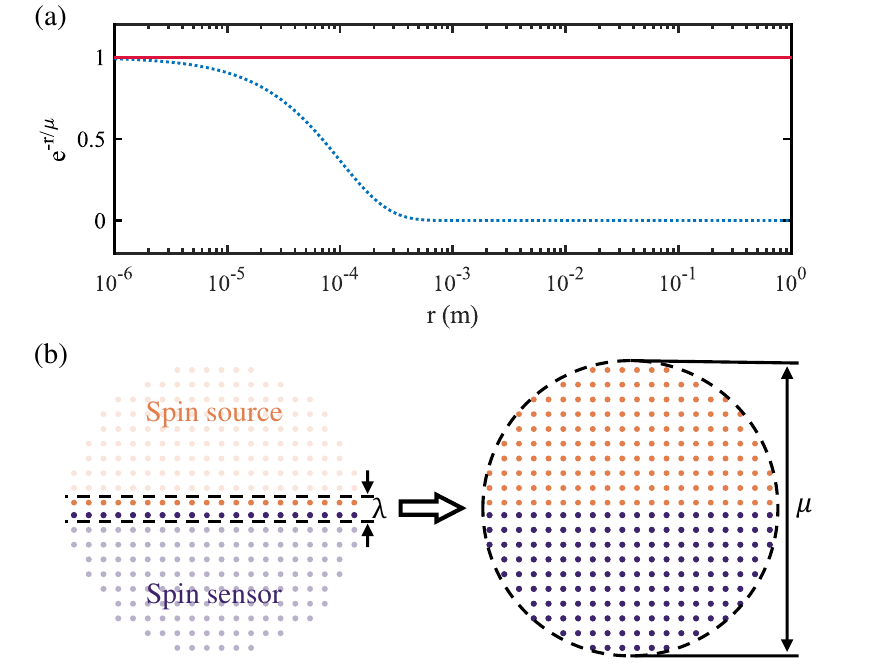}
  \caption{ 
      (a) The distance dependence of the exponential factor $e^{-r/\mu}$ at $t=0$ with
          $\gamma\omega = m_{\rm X}$ (blue dotted line) and $\gamma\omega = 0$ (red solid line).
          Here $m_{\rm X}$ is set to 2 meV, corresponding to a force range of $\lambda = 0.1$ mm.
      (b) An illustration of the concept of the near-threshold enhancement.
          At the static scenario, interacting fermions are limited to a thin layer with the thickness $\lambda$,
            while at the near-threshold scenario, all fermions within the distance $\mu$, 
            which is orders of magnitude larger than $\lambda$, can contribute to the total interaction signal.
  }
  \label{Figure2}
\end{figure}

Now we would like to extend the problem
  by assuming that the spin of fermion 1 is oscillating at a certain frequency $\omega$,
  $\bm{\sigma}_{1} \rightarrow \bm{\sigma}_{1}e^{-i\omega x^{0}}$,
  as shown in Fig. \ref{Figure1}(b).
Then Eq. \ref{equ4} is modified to
\begin{equation}
  \begin{aligned}
    \langle iT_{\rm AA}\rangle = -ig_{\rm A}^{\psi_{1}}g_{\rm A}^{\psi_{2}}4m_{\psi_{1}}m_{\psi_{2}}
                                  \frac{\mathcal{O}_{2}+\frac{M^{2}}{m_{\rm X}^{2}}\mathcal{O}_{3}}{q^{2}-m_{\rm X}^{2}}\\
                                  \times(2\pi )^{4}\delta^{4}(p'+k'-p-k-\Omega),
    \label{equ5}
  \end{aligned}
\end{equation} 
  where $\Omega = (\omega,\bm{0})^{\rm T}$ is the injected 4-momentum in the lab frame $\mathcal{K}$,
  and $q = p'-p-\Omega$.
The expression of the scattering amplitude includes two parts:
  the first part gives the potential energy in the momentum space,
\begin{equation}
  V_{\rm AA}(q) = -g_{\rm A}^{\psi_{1}}g_{\rm A}^{\psi_{2}}(\mathcal{O}_{2}+\frac{M^{2}}{m_{\rm X}^{2}}\mathcal{O}_{3})\frac{1}{q^{2}-m_{\rm X}^{2}},
\end{equation} 
  while the remaining part reflects the conservation law of energy and momentum.

Finding the expression of the potential in the position space requires inverting the Fourier transformation,
  i.e., $V_{\rm AA}(x) = \int V_{\rm AA}(p) e^{iq\cdot x} d^{3}\bm{q}/(2\pi)^{3}$.
Without losing generality, 
  a frame $\widetilde{\mathcal{K}}$ where $\widetilde{\bm{p}}+\widetilde{\bm{p}}'=0$ is chosen for convenience, 
  and thus $\widetilde{p}'^{0}=\widetilde{p}^{0}$.
In such a frame, 
  the injected momentum undergoes a Lorentz transformation, 
  giving $\widetilde{\Omega} = (\omega\gamma,\bm{\alpha})$ 
  and $\widetilde{q} = (-\omega\gamma,\widetilde{\bm{p}}'-\widetilde{\bm{p}}-\bm{\alpha})$,
  where $\gamma = (1-|\bm{\beta}|^{2})^{-1}$ is the Lorentz factor, $\bm{\beta}$ is the relative velocity between $\widetilde{\mathcal{K}}$ and $\mathcal{K}$,
  and $\bm{\alpha} = -\omega\gamma\bm{\beta}$.
After integrating $\widetilde{\bm{q}}$ over the whole momentum space, we get the expression for the potential between two fermions in the position space,
\begin{equation}
  V_{\rm AA} = -g_{\rm A}^{\psi_{1}}g_{\rm A}^{\psi_{2}}(\mathcal{V}_{2}+\frac{M^{2}}{m_{\rm X}^{2}}\mathcal{V}_{3})
               e^{-\widetilde{r}\sqrt{m_{\rm X}^{2}-\omega^{2}\gamma^{2}}}e^{-i\omega\gamma\widetilde{t}}
\end{equation} 
By introducing an inverse Lorentz transformation, we can finally get the expression of the potential in the lab frame.
In the nonrelativistic limit, the potential is
\begin{equation}
  V_{\rm AA} \approx -g_{\rm A}^{\psi_{1}}g_{\rm A}^{\psi_{2}}(\mathcal{V}_{2}+\frac{M^{2}}{m_{\rm X}^{2}}\mathcal{V}_{3})
               e^{-r/\mu}e^{-i\omega t}
  \label{Eq8}
\end{equation} 
  where $\mu^{-2} = m_{\rm X}^{2}-\omega^{2}\gamma^{2}$ is the effective force range.
This result can be further extended to other types of spin-dependent exotic interactions such as $V_{\rm AV},V_{\rm VV},V_{\rm pp},$ and so on,
  by replacing $\mathcal{O}_{2,3}$ by other scalar operators $\mathcal{O}_{i}(i=2,\dots,16)$ in the derivation
  (see more details in Sec. \uppercase\expandafter{\romannumeral3} of Supplemental Material \cite{spp}).
However, this result do not apply to $V_{1}$ since $V_{1}$ is spin-independent.

Equation \ref{Eq8} indicates that an oscillating spin gives rise to exotic interactions that are quite different from the static version.
A natural consequence it that the exotic interactions oscillate at the same frequency as the source.
A more fascinating outcome is that the effective force range $\mu$ becomes larger as $\omega$ increases.
When the frequency matches the mass of the new bosons, i.e., $\gamma\omega = m_{\rm X}$,
  the exponential decay of the exotic interactions completely vanishes,
  just like the new bosons had no mass at all.
At $t=0$, the distance dependence of the exponential factor when $\gamma\omega = m_{\rm X}$ and $\gamma\omega = 0$
  are shown as the red solid line and the blue dotted line in Fig. \ref{Figure2} (a), respectively.
It is noteworthy that in practice an infinite $\mu$ is unreachable, 
  since (1) $\gamma$ doesn't have an explicit value due to the thermal motion of fermions;
  (2) the oscillation frequency of the spins is not perfectly pure-tone
  (see more details in Sec. \uppercase\expandafter{\romannumeral4} of Supplemental Material \cite{spp}).
However, in the near-threshold regime, $\mu$ can be many orders larger than $\lambda$.
As shown in Fig. \ref{Figure2}(b), at the static scenario only fermions within the force range $\lambda$ can contribute to the exotic interactions,
  while at the near-threshold scenario even the fermions at distances larger than $\lambda$ (but smaller than  $\mu$) can also contribute to the exotic interactions.
Consequently, the sizes of the interacting objects are no longer limited by the wavelength of the new bosons,
  and both the signal of the exotic interactions and the sensitivity can be greatly enhanced by improving the number of fermions.

\begin{figure}[b]
  \centering
  \includegraphics[width=1\columnwidth]{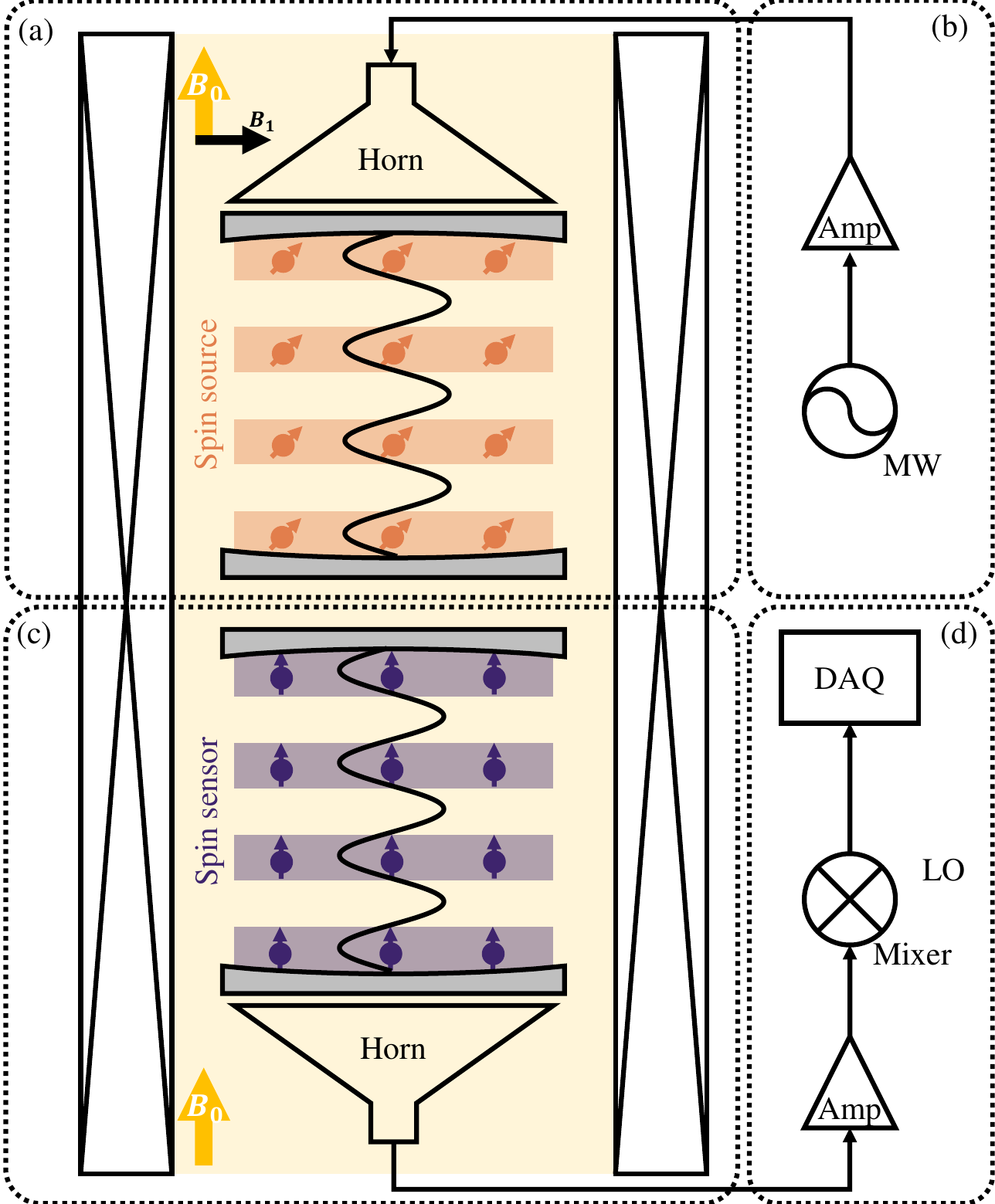}
  \caption{ 
      Schematic diagram of the proposed experiment.
      (a) The spin source.
          YIG bulks (orange) are driven by a pure-tone microwave to generate an oscillating exotic interaction. 
          They are cut into slices and equally spaced in a resonant cavity (gray), so that all spins contribute positively to the exotic interaction.
      (b) The driving system.
          A microwave is generated, amplified, and fed into the source cavity.
      (c) The spin sensor.
          YIG bulks (purple) in another cavity (gray) perform as the sensor.
          When the spins sense the exotic interaction, they will precess and radiate microwave photons.
          Both the spin source and the spin sensor are placed in a homogeneous magnetic field $B_{0}$, as the yellow area shows.
      (d) The readout system.
          The microwave signal from the sensor cavity is amplified, down-converted, and finally collected by a data acquisition card.
  }
  \label{Figure3}
\end{figure}
\begin{figure}
  \centering
  \includegraphics[width=1\columnwidth]{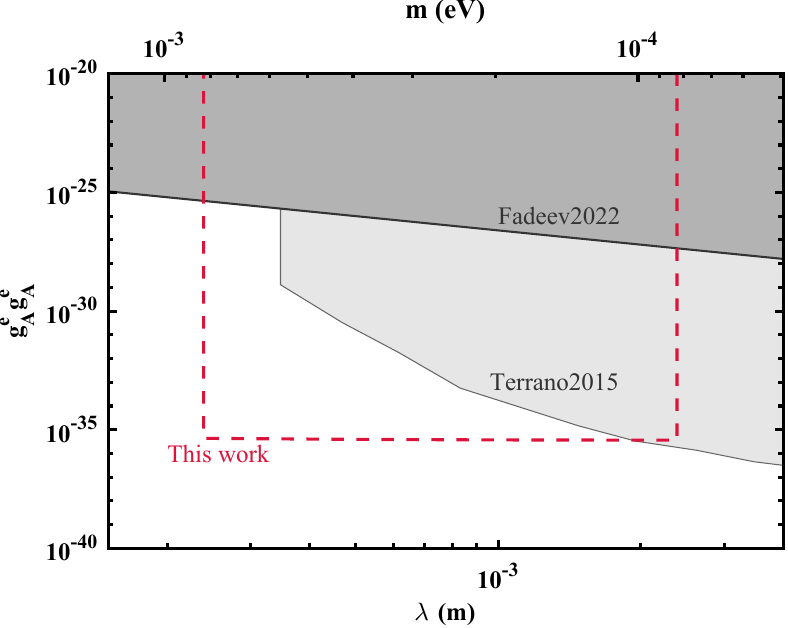}
  \caption{ 
      Expected constraints on $g_{\rm A}^{\rm e}g_{\rm A}^{\rm e}$ as the driving frequencies are swept from 20 GHz to 200 GHz.
        At 800 $\mu$eV the expected constraint surpasses existing results by 10 orders of magnitude.
  }
  \label{Figure4}
\end{figure}

Based on this theory, we propose a method for searching for exotic interactions taking advantage of the near-threshold enhancement.
The schematic of the proposed experimental setup is shown in Fig. \ref{Figure3}.
Four parts are included in the setup: (a) a spin source, (b) a driving system, (c) a spin sensor and (d) a readout system.
The spin source is positioned in an external magnetic field $\bm{B_{0}} = B_{0}\bm{\widehat{z}}$, 
  so its resonant frequency is $\omega_{0} = \gamma_{\rm e}B_{0}$,
  where $\gamma_{\rm e}$ is the gyromagnetic ratio.
When the spin source is driven by a continuous resonant microwave, the spins will deviate from and precess around the z-axis.
At low-excited limit, the transverse and longitudinal components of the source spins in the lab frame are
  $S_{\bot} = S_{\rm z}^{0}\gamma_{\rm e}B_{1}T_{2}^{*}e^{-i\omega_{0}t}$ and $S_{\rm z} = S_{\rm z}^{0}$, respectively,
  where $B_{1}$ is the amplitude of the magnetic component of the driving microwave field, 
  $T_{2}^{*}$ is the transverse relaxation time,
  and $S_{\rm z}^{0}$ is the initial value of $S_{\rm z}$.
For a polarized spin-$1/2$ system, $S_{\rm z}^{0} = 1/2$.
Therefore, the transverse component of the spins can generate an oscillating exotic interaction potential, which is effective to a pseudomagnetic field and has a frequency $\omega_{0}$.
Since the sensor spins are also positioned in the external magnetic field $\bm{B_{0}}$ and thus have the resonant frequency $\omega_{0}$,
  they will be resonantly driven by the pseudomagnetic field and radiate microwaves at the frequency $\omega_{0}$.
The power of the radiation is 
  $P = \gamma_{\rm e}\mu_{\rm B}N\omega_{0} |\bm{B}_{\rm exo}|^{2}T_{2}^{*}$
  where $\mu_{\rm B}$ is the Bohr magneton, $N$ is the total number of the spins in the sensor, 
  $T_{2}^{*}$ is the transverse relaxation time of the spins,
  and $\bm{B}_{\rm exo}$ is the pseudomagnetic field induced by the exotic interactions \cite{2020Crescini}.
For the pseudoscalar/pseudoscalar coupling, $\bm{B}_{\rm exo}$ is defined as 
  $V_{\rm pp} = \gamma_{\rm e}\bm{B}_{\rm exo}\cdot\bm{\sigma}_{2}$, 
  where $\bm{\sigma}_{2}$ refers to a sensor spin \cite{2021Jiao}.
Both the spin source and the spin sensor are placed in resonant cavities, so that microwave with the frequency $\omega_{0}$ can form standing waves.
The radiated microwave is amplified by a low-noise-amplifier, down-converted by a mixer, and finally collected by a data acquisition card.

The experimental parameters are carefully designed to optimize the sensitivity of the setup to exotic interactions.
Here we focus on a well-motivated mass range of the new bosons, 80$\sim$800 $\mu$eV, 
  corresponding to the frequency range from  20 GHz to 200 GHz \cite{2015Kawasaki,2016Borsanyi}.
To achieve such resonant frequencies, an external magnetic field $B_{0}$ of 0.7$\sim$7 T is required.
Yttrium iron garnet (YIG) is chosen as both the spin source and the spin sensor for its high spin density ($n_{\rm YIG}=2\times 10^{22}\ \rm cm^{-3}$),
  high polarization (nearly fully polarized),
  and relatively long relaxation time ($T_{2}^{*} \approx 100$ ns).
Both the spin source and the spin sensor are assumed to be cylinders with a radius of $R=5$ cm and a height of $H=10$ cm.
As the size of the spin source and the spin sensor exceeds the wavelength of the microwave with the frequency $\omega_{0}$,
  the source and the sensor should be made into multilayer structures to avoid coherence cancellation of the signal.
The thickness of each layer and their distance are half of the wavelength of the microwave, 
  so that the microwave field sensed by each electron spin in the YIG bulks can make positive contributions to driving the spins.
It is noteworthy that, although the phase of the microwave in a YIG bulk may still have $o(1)$ variations,
  YIG is resistant to the phase variations at such level. 
This is because YIG is a kind of ferrimagnetic material, whose electron spins strongly couple to each other and thus always precess with the same phase.
A coherent microwave source with a linewidth $\delta\omega_{0} = \omega_{0}/Q$ and a power of $P_{\rm in} = 1$ W is used to drive the spin source,
  where $Q = 5\times 10^{7}$ is the quality factor of the microwave source.
Although it is difficult to sweep through the total frequency range with a single setup,
  the frequency range can be covered by merely replacing the microwave devices with different operating bands.
The whole setup is cooled down to a temperature of $T=4$ K.

The constraints on $g_{\rm A}^{\rm e}g_{\rm A}^{\rm e}$ depends on the noise level of the readout system.
For linear amplifiers, the standard quantum limit with a noise spectral density of $S=\omega_{0}$ is achievable \cite{1982Caves}.
As the driving frequencies are swept from 20 GHz to 200 GHz, 
  the expected constraints are shown as the red line in Fig. \ref{Figure4}.
The integration time is 1 seconds for each frequency and thus 3 years for the total frequency range.
The gray regions stand for the parameter spaces excluded by previous studies \cite{2015Terrano,2017Ficek,2022Fadeev}.
At $m_{\rm X}=800\ \mu$eV, the expected constraints on $g_{\rm A}^{\rm e}g_{\rm A}^{\rm e}$ 
  of our method exceeds the previous results by around 10 orders of magnitude.
  
To conclude, we studied the near-threshold enhancement of exotic interactions, and proposed a method to search for exotic interactions through it.
When the oscillation frequency of fermions matches the mass of the new bosons,
  the effective force range is magnified, 
  and the exotic interaction is enhanced.
Consequently, by taking advantage of the near-threshold enhancement, 
  experimental searching for exotic interactions on small scales can be enhanced since the limitation from the wavelength of the new bosons is broken.
At 800 $\mu$eV, we expect an improvement on the constraints on the pseudovector/pseudovector coupling of around 10 orders of magnitude.
The concept of near-threshold searches proposed in this work can also be adapted to a broad class of physical systems --
  such as SERF magnetometers, NV centers, and mechanical cantilevers --
  and be further extended to explore other types of interactions.
This work paves a way for searching for new bosons beyond the Standard Model and will boost the study on new physics on small length scales.
%

The authors are grateful to Prof. Renbao Liu for the valuable discussions. 
This work was supported by NSFC (T2388102, 12150010, 12261160569),
  the Innovation Program for Quantum Science and Technology (2021ZD0302200).
X.R. is thankful for the support by the Major Frontier Research Project of the University of Science and Technology of China (Grant No. LS9990000002).  

\appendix

\section{Scalar operators and exotic interactions\label{Sec1}}

In the two-fermion scattering reaction shown in Fig. 1 in the main text, 
16 scalar operators can be constructed:

\begin{subequations}
  \begin{equation} 
      \mathcal{O}_{1}     = 1,
  \end{equation}
  \begin{equation} 
      \mathcal{O}_{2}     = \bm{\sigma}_{1}\cdot\bm{\sigma}_{2},
  \end{equation}
  \begin{equation} 
      \mathcal{O}_{3}     = \frac{1}{M^{2}}(\bm{\sigma}_{1}\cdot\bm{q})(\bm{\sigma}_{2}\cdot\bm{q}),
  \end{equation}
  \begin{equation} 
      \mathcal{O}_{4,5}   = \frac{i}{2M^{2}}(\bm{\sigma}_{1}\pm\bm{\sigma}_{2})\cdot(\bm{P}\times\bm{q}),
  \end{equation}
  \begin{equation} 
      \mathcal{O}_{6,7}   = \frac{i}{2M^{2}}[(\bm{\sigma}_{1}\cdot\bm{P})(\bm{\sigma}_{2}\cdot\bm{q})\pm(\bm{\sigma}_{2}\cdot\bm{P})(\bm{\sigma}_{1}\cdot\bm{q})],
  \end{equation}
  \begin{equation} 
      \mathcal{O}_{8}     = \frac{1}{M}(\bm{\sigma}_{1}\cdot\bm{P})(\bm{\sigma}_{2}\cdot\bm{P}),
  \end{equation}
  \begin{equation} 
      \mathcal{O}_{9,10}  = -\frac{i}{2M}(\bm{\sigma}_{1}\pm\bm{\sigma}_{2})\cdot\bm{q},
  \end{equation}
  \begin{equation} 
      \mathcal{O}_{11}    = \frac{i}{M}(\bm{\sigma}_{1}\times\bm{\sigma}_{2})\cdot\bm{q},
  \end{equation}
  \begin{equation} 
      \mathcal{O}_{12,13} = \frac{1}{2M}(\bm{\sigma}_{1}\pm\bm{\sigma}_{2})\cdot\bm{P},
  \end{equation}
  \begin{equation} 
      \mathcal{O}_{14}    = \frac{1}{M}(\bm{\sigma}_{1}\times\bm{\sigma}_{2})\cdot\bm{P},
  \end{equation}
  \begin{equation} 
   \begin{aligned}
      \mathcal{O}_{15}    = \frac{1}{2M^{3}}\{[\bm{\sigma}_{1}\cdot(\bm{P}\times\bm{q})](\bm{\sigma}_{2}\cdot\bm{q})\\
      +[\bm{\sigma}_{2}\cdot(\bm{P}\times\bm{q})](\bm{\sigma}_{1}\cdot\bm{q})\},
   \end{aligned}
  \end{equation}
  \begin{equation} 
  \begin{aligned}
      \mathcal{O}_{16}    = \frac{i}{2M^{3}}\{[\bm{\sigma}_{1}\cdot(\bm{P}\times\bm{q})](\bm{\sigma}_{2}\cdot\bm{P})\\
      +[\bm{\sigma}_{2}\cdot(\bm{P}\times\bm{q})](\bm{\sigma}_{1}\cdot\bm{P})\},
      \end{aligned}
  \end{equation}
\end{subequations}
  where $M = m_{\psi_{1}}m_{\psi_{2}}/(m_{\psi_{1}}+m_{\psi_{2}})$ is the reduced mass of the two-fermion system.
Any other scalars related to $\bm{\sigma}_{1}$, $\bm{\sigma}_{2}$, $\bm{q}$, 
  and $\bm{P}$ and be expressed as a linear combination of these operators \cite{2006Dobrescu}.

From the these scalar operators,
  16 interactions with the following form can be derived:
\begin{equation}
  V_{\rm i} = f_{\rm i}\mathcal{V}_{\rm i}e^{-m_{\rm X}r}(i=1,2,\dots 16),
\end{equation}
  where $f_{\rm i}$ is the coupling strength, $r$ is the distance between the two fermions, $m_{\rm X}$ is the mass of the new bosons, and

\begin{subequations}
  \begin{equation} 
      \mathcal{V}_{1}     = \frac{1}{4\pi r},
      \label{V1}
  \end{equation}
  \begin{equation} 
      \mathcal{V}_{2}     = \frac{1}{4\pi r}\bm{\sigma}_{1}\cdot\bm{\sigma}_{2},
      \label{V2}
  \end{equation}
  \begin{equation} 
    \begin{aligned}
        \mathcal{V}_{3}   =& \frac{1}{4\pi M^{2}r^{3}}[\bm{\sigma}_{1}\cdot\bm{\sigma}_{2}(1+\frac{r}{\lambda})\\
          &-(\bm{\sigma}_{1}\cdot\bm{\widehat{r}})(\bm{\sigma}_{2}\cdot\bm{\widehat{r}})(3+\frac{3r}{\lambda}+\frac{r^{2}}{\lambda^{2}})],
    \end{aligned}
    \label{V3}
  \end{equation}
  \begin{equation} 
      \mathcal{V}_{4,5}   = -\frac{1}{8\pi Mr^{2}} (\bm{\sigma}_{1}\pm\bm{\sigma}_{2})\cdot(\bm{\beta}\times\bm{\widehat{r}})                 (1+\frac{r}{\lambda}),
      \label{V45}
  \end{equation}
  \begin{equation} 
    \begin{aligned}
      \mathcal{V}_{6,7}   = -&\frac{1}{8\pi M^{2}r^{2}} [(\bm{\sigma}_{1}\cdot\bm{\beta})(\bm{\sigma}_{2}\cdot\bm{\widehat{r}})\\
        &\pm(\bm{\sigma}_{2}\cdot\bm{\beta})(\bm{\sigma}_{1}\cdot\bm{\widehat{r}})] (1+\frac{r}{\lambda}),
      \label{V67}
    \end{aligned}
  \end{equation}
  \begin{equation} 
      \mathcal{V}_{8}     = \frac{1}{4\pi r}(\bm{\sigma}_{1}\cdot\bm{\beta})(\bm{\sigma}_{2}\cdot\bm{\beta}),
      \label{V8}
  \end{equation}
  \begin{equation} 
      \mathcal{V}_{9,10}  = -\frac{1}{8\pi Mr^{2}} (\bm{\sigma}_{1}\pm\bm{\sigma}_{2})\cdot\bm{\widehat{r}} (1+\frac{r}{\lambda}),
      \label{V910}
  \end{equation}
  \begin{equation} 
      \mathcal{V}_{11}    = -\frac{1}{4\pi Mr^{2}} (\bm{\sigma}_{1}\times\bm{\sigma}_{2})\cdot\bm{\widehat{r}} (1+\frac{r}{\lambda}),
      \label{V11}
  \end{equation}
  \begin{equation} 
      \mathcal{V}_{12,13} = \frac{1}{8\pi r}(\bm{\sigma}_{1}\pm\bm{\sigma}_{2})\cdot\bm{\beta},
      \label{V1213}
  \end{equation}
  \begin{equation} 
      \mathcal{V}_{14}    = \frac{1}{4\pi r}(\bm{\sigma}_{1}\times\bm{\sigma}_{2})\cdot\bm{\beta},
      \label{V14}
  \end{equation}
  \begin{equation} 
  \begin{aligned}
      \mathcal{V}_{15}    = \frac{-1}{8\pi M^{2}r^{3}}\{[\bm{\sigma}_{1}\cdot(\bm{\widehat{r}}\times\bm{\beta})](\bm{\sigma}_{2}\cdot\bm{\widehat{r}})\\
      +[\bm{\sigma}_{2}\cdot(\bm{\widehat{r}}\times\bm{\beta})](\bm{\sigma}_{1}\cdot\bm{\widehat{r}})\}(3+\frac{3r}{\lambda}+\frac{r^{2}}{\lambda^{2}}),
      \label{V15}
      \end{aligned}
  \end{equation}
  \begin{equation} 
  \begin{aligned}
      \mathcal{V}_{16}    = \frac{-1}{8\pi Mr^{2}}\{[\bm{\sigma}_{1}\cdot(\bm{\widehat{r}}\times\bm{\beta})](\bm{\sigma}_{2}\cdot\bm{\beta})\\
      +[\bm{\sigma}_{2}\cdot(\bm{\widehat{r}}\times\bm{\beta})](\bm{\sigma}_{1}\cdot\bm{\beta})\}(1+\frac{r}{\lambda}),
      \label{V16}
      \end{aligned}
  \end{equation}
  \label{V}
\end{subequations}
  where $\bm{\beta}$ is the relative velocity between the two fermions \cite{2006Dobrescu}.

In this paper, we introduce two new scalar operators
\begin{subequations}
  \begin{equation} 
      \mathcal{O}_{14q+15} = \frac{1}{M^{3}}(\bm{\sigma}_{1}\cdot\bm{q})[\bm{\sigma}_{2}\cdot(\bm{P}\times\bm{q})],
  \end{equation}
  \begin{equation} 
      \mathcal{O}_{11p+16} = \frac{i}{M^{3}}(\bm{\sigma}_{1}\cdot\bm{P})[\bm{\sigma}_{2}\cdot(\bm{P}\times\bm{q})].
  \end{equation}
\end{subequations}
The corresponding interactions are
\begin{subequations}
  \begin{equation} 
    \begin{aligned}
      \mathcal{V}_{14q+15}  &= \frac{1}{4\pi M^{2}r^{3}}[(\bm{\sigma}_{1}\times\bm{\sigma}_{2})\cdot\bm{\beta}(1+\frac{r}{\lambda})\\
                            &-(\bm{\sigma}_{1}\cdot\bm{\widehat{r}})(\bm{\sigma}_{2}\times\bm{\beta})\cdot\bm{\widehat{r}}(3+\frac{3r}{\lambda}+\frac{r^{2}}{\lambda^{2}})],
      \label{V15}
    \end{aligned}
  \end{equation}
  \begin{equation} 
    \begin{aligned}
      \mathcal{V}_{11p+16}  &= -\frac{1}{4\pi Mr^{2}}(\bm{\sigma}_{1}\times\bm{\beta})(\bm{\sigma}_{2}\times\bm{\beta})\cdot\bm{\widehat{r}}(1+\frac{r}{\lambda}),\\
      \label{V16}
    \end{aligned}
\end{equation}
\label{V}
\end{subequations}
respectively. These two terms are also discussed in \cite{2025Cong}.

\section{Exotic interactions respect to the new bosons and the coupling}

New bosons beyond the Standard Model are predicted to mediate exotic interactions, which can be expressed using Eq. \ref{V1}, \ref{V2}, \ref{V3}, \ref{V45}, \ref{V67}, \ref{V8}, \ref{V910}, \ref{V11}, \ref{V1213}, \ref{V14}, \ref{V15}, and \ref{V16}.

The exotic interactions induced by axion/axionlike particles are
\begin{subequations}
  \begin{equation}
    V_{\rm pp} = g_{\rm p}^{1}g_{\rm p}^{2}\frac{M^{2}}{4m_{\psi_{1}}m_{\psi_{2}}}\mathcal{V}_{3}e^{-r/\lambda},
  \end{equation}
  \begin{equation}
    V_{\rm ps} = g_{\rm p}^{1}g_{\rm s}^{2}\frac{M}{2m_{\psi_{1}}}(\mathcal{V}_{9}+\mathcal{V}_{10})e^{-r/\lambda},
  \end{equation}
  \begin{equation}
    V_{\rm ss} = -g_{\rm s}^{1}g_{\rm s}^{2}\mathcal{V}_{1}e^{-r/\lambda},
  \end{equation}
\end{subequations}
the exotic interactions induced by $Z'$ bosons are
\begin{subequations}
  \begin{equation}
    V_{\rm AA} = -g_{\rm A}^{1}g_{\rm A}^{2}(\mathcal{V}_{2}+\frac{M^{2}}{m_{\rm X}^{2}}\mathcal{V}_{3})e^{-r/\lambda},
  \end{equation}
  \begin{equation}
    V_{\rm AV} = g_{\rm A}^{1}g_{\rm V}^{2}(\frac{M}{2m_{2}}\mathcal{V}_{11}+\mathcal{V}_{12}+\mathcal{V}_{13})e^{-r/\lambda},
  \end{equation}
  \begin{equation}
    V_{\rm VV} = g_{\rm V}^{1}g_{\rm V}^{2}(\mathcal{V}_{1}+\frac{m_{\rm X}^{2}\mathcal{V}_{2}+M^{2}\mathcal{V}_{3}}{4m_{\psi_{1}}m_{\psi_{2}}})e^{-r/\lambda},
  \end{equation}
\end{subequations}
and the exotic interactions induced by paraphotons are
\begin{subequations}
  \begin{equation}
    \begin{aligned}
      V_{\rm TT} &= \frac{4\textrm{Re}(C_{\psi 1})\textrm{Re}(C_{\psi 2})v_{h}^{2}}{\Lambda^{4}}\\
                  &(\frac{m_{\rm X}^{4}}{4m_{\psi_{1}}m_{\psi_{2}}}\mathcal{V}_{1}+m_{\rm X}^{2}\mathcal{V}_{2}+M^{2}\mathcal{V}_{3})e^{-r/\lambda},
    \end{aligned}
  \end{equation}
  \begin{equation}
    \begin{aligned}
      V_{\rm \widetilde{T}T} &= \frac{4\textrm{Im}(C_{\psi 1})\textrm{Re}(C_{\psi 2})v_{h}^{2}}{\Lambda^{4}}\\
                            &[\frac{Mm_{\rm X}^{2}}{2m_{\psi_{2}}}(\mathcal{V}_{9}+\mathcal{V}_{10})+M^{2}\mathcal{V}_{14p+15}]e^{-r/\lambda},
    \end{aligned}
  \end{equation}
  \begin{equation}
    V_{\rm \widetilde{T}\widetilde{T}} = 4\textrm{Im}(C_{\psi 1})\textrm{Im}(C_{\psi 2})\frac{v_{h}^{2}}{\Lambda^{4}}
                                          M^{2}\mathcal{V}_{3}e^{-r/\lambda}.
  \end{equation}
\end{subequations}
Note that in this paper we only focus on the non-relativistic limit, 
  so the high-order terms $\mathcal{V}_{4,5}$, $\mathcal{V}_{8}$, and $\mathcal{V}_{16}$ are neglected here.
More details of these terms can be found in \cite{2025Cong}.

\begin{widetext}
\section{Near-threshold enhancement of exotic interactions}
\subsection{Detailed derivation}

\begin{figure}
  \centering
  \includegraphics[width=0.5\columnwidth]{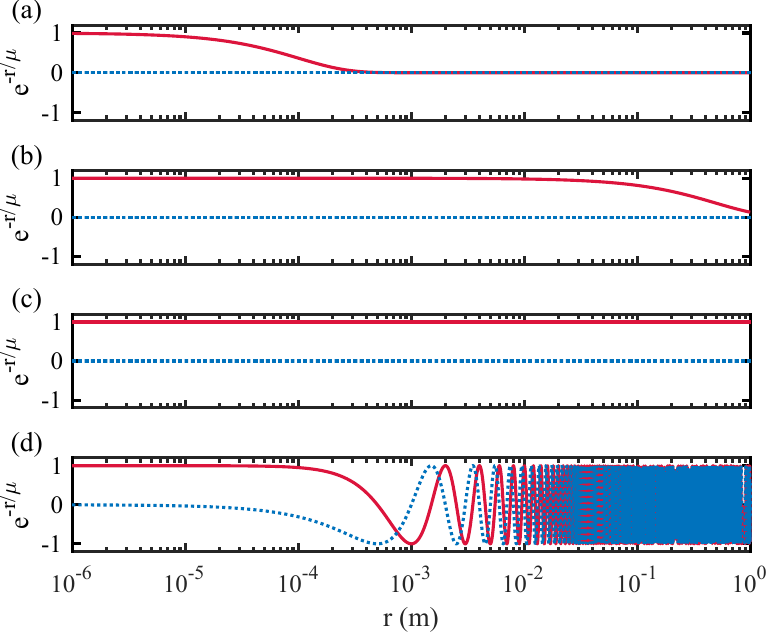}
  \caption{ 
      The distance dependence of the exponential factor $e^{-r/\mu}$ at $t=0$ with $m_{\rm X} = 2\ \rm meV$ and
      (a) $\omega = 0$;
      (b) $0<\gamma\omega<m_{\rm X}$;
      (c) $\gamma\omega = m_{\rm X}$;
      (d) $\gamma\omega > m_{\rm X}$.
      The red solid lines refer to the real part while the blue dotted lines refer to the imaginary part.
  }
  \label{FigureS1}
\end{figure}

In this subsection we provide more details in the proof of the near-threshold enhancement.
The pseudovector/pseudovector term in Eq.3 in the main text is
\begin{equation}
  \begin{aligned}
  \langle iT_{\rm AA}\rangle &=    
      -g_{\rm A}^{\psi_{1}}g_{\rm A}^{\psi_{2}}
      \contraction[1ex]{\bra{\bm{p}'\bm{k}'}\int d^{4}x }                                                                                                 {Z'_{\mu}}{\bar{\psi}\gamma^{\mu}\gamma^{5}\psi \int d^{4}y}                                      {Z'_{\nu}}
      \contraction[2ex]{\bra{}}                                                                                                                           {\bm{p}'}   {\bm{k}'\int d^{4}x Z'_{\mu}}                                                  {\bar{\psi}}
      \contraction[2ex]{\bra{\bm{p}'\bm{k}'}\int d^{4}x Z'_{\mu}\bar{\psi}\gamma^{\mu}\gamma^{5}}                                                         {\psi}      {\int d^{4}y Z'_{\nu}\bar{\psi}\gamma^{\nu}\gamma^{5}\psi}                  {\bm{p}}
      \contraction[3ex]{\bra{\bm{p}'}}                                                                                                                    {\bm{k}'}   {\int d^{4}x Z'_{\mu}\bar{\psi}\gamma^{\mu}\gamma^{5}\psi \int d^{4}y Z'_{\nu}}{\bar{\psi}}
      \contraction[3ex]{\bra{\bm{p}'\bm{k}'}\int d^{4}x Z'_{\mu}\bar{\psi}\gamma^{\mu}\gamma^{5}\psi \int d^{4}y Z'_{\nu}\bar{\psi}\gamma^{\nu}\gamma^{5}}{\psi}      {\bm{p}}                                                                         {\bm{k}}
      \bra{\bm{p}'\bm{k}'}\int d^{4}x Z'_{\mu}\bar{\psi}\gamma^{\mu}\gamma^{5}\psi \int d^{4}y Z'_{\nu}\bar{\psi}\gamma^{\nu}\gamma^{5}\psi\ket{\bm{p}\bm{k}}\\
                            &= -g_{\rm A}^{\psi_{1}}g_{\rm A}^{\psi_{2}}\int d^{4}x\int d^{4}y\int\frac{d^{4}q}{(2\pi)^{4}}
                            [\bar{u}(p')\gamma^{\mu}\gamma^{5}u(p)]\mathcal{D}_{\mu\nu}[\bar{u}(k')\gamma^{\nu}\gamma^{5}u(k)]e^{i(p'-p-q)\cdot x + i(k'-k+q)\cdot y},
  \end{aligned}
  \label{Eq9}
\end{equation} 
  where 
  $u(l) = (\sqrt{l\cdot\sigma},\sqrt{l\cdot\bar{\sigma}})$, 
  $\bar{u}(l) = u^{\dagger}(l)\gamma^{0}$, 
  $\sigma = (1,\bm{\sigma})$, $\bar{\sigma} = (1,-\bm{\sigma})$, $l$ is an arbitrary 4-momentum,
  and $\mathcal{D}_{\mu\nu}$ is the propagator of a massive spin-1 boson.
In the nonrelativistic limit, the spinor products and the propagator are
\begin{subequations}
  \begin{equation}
    \bar{u}(p')\gamma^{\mu}\gamma^{5}u(p) \approx (0,2m_{\psi_{1}}\bm{\sigma}_{1}),
  \end{equation}
  \begin{equation}
    \bar{u}(k')\gamma^{\nu}\gamma^{5}u(k) \approx (0,2m_{\psi_{2}}\bm{\sigma}_{2}),
  \end{equation}
  \begin{equation}
    \mathcal{D}_{\mu\nu} = -i\frac{g_{\mu\nu}-q_{\mu}q_{\nu}}{q^{2}-m_{\rm X}^{2}}.
  \end{equation}
  \label{Eq10}
\end{subequations}
Therefore, the pseudovector/pseudovector interaction element is
\begin{equation}
  \begin{aligned}
    \langle iT_{\rm AA}\rangle  = -g_{\rm A}^{\psi_{1}}g_{\rm A}^{\psi_{2}}4m_{\psi_{1}}m_{\psi_{2}}\int d^{4}x\int d^{4}y\int\frac{d^{4}q}{(2\pi)^{4}}
                  \frac{i[(\bm{\sigma}_{1}\cdot\bm{\sigma}_{2})+(\bm{\sigma}_{1}\cdot q)(\bm{\sigma}_{2}\cdot q)/m_{\rm X}^{2}]}{q^{2}-m_{\rm X}^{2}}e^{i(p'-p-q)\cdot x + i(k'-k+q)\cdot y}.
    \label{equ11}
  \end{aligned}
\end{equation} 
By replacing $\bm{\sigma}_{1}$ with $\bm{\sigma}_{1}e^{-i\omega x^{0}}$, this expression can be modified into 
\begin{equation}
  \begin{aligned}
    \langle iT_{\rm AA}\rangle  = -g_{\rm A}^{\psi_{1}}g_{\rm A}^{\psi_{2}}4m_{\psi_{1}}m_{\psi_{2}}\int d^{4}x\int d^{4}y\int\frac{d^{4}q}{(2\pi)^{4}}
                  \frac{i[(\bm{\sigma}_{1}\cdot\bm{\sigma}_{2})+(\bm{\sigma}_{1}\cdot q)(\bm{\sigma}_{2}\cdot q)/m_{\rm X}^{2}]}{q^{2}-m_{\rm X}^{2}}
                  e^{i(p'-p-q-\Omega)\cdot x + i(k'-k+q)\cdot y},
    \label{equ7}
  \end{aligned}
\end{equation} 
  i.e.
\begin{equation}
  \begin{aligned}
    \langle iT_{\rm AA}\rangle  = -g_{\rm A}^{\psi_{1}}g_{\rm A}^{\psi_{2}}4m_{\psi_{1}}m_{\psi_{2}}\int d^{4}x\int d^{4}y\int\frac{d^{4}q}{(2\pi)^{4}}
                  \frac{i(\mathcal{O}_{2}+\frac{M^{2}}{m_{\rm X}^{2}}\mathcal{O}_{3})}{q^{2}-m_{\rm X}^{2}}
                  e^{i(p'-p-q-\Omega)\cdot x + i(k'-k+q)\cdot y}.
    \label{equ7}
  \end{aligned}
\end{equation} 
Integrating over $x$, $y$ and $q$ gives
\begin{equation}
  \begin{aligned}
    \langle iT_{\rm AA}\rangle = -ig_{\rm A}^{\psi_{1}}g_{\rm A}^{\psi_{2}}4m_{\psi_{1}}m_{\psi_{2}}
                                  \frac{\mathcal{O}_{2}+\frac{M^{2}}{m_{\rm X}^{2}}\mathcal{O}_{3}}{q^{2}-m_{\rm X}^{2}}
                                  \times(2\pi )^{4}\delta^{4}(p'+k'-p-k-\Omega),
    \label{equ5}
  \end{aligned}
\end{equation} 
  where $\Omega = (\omega,\bm{0})^{\rm T}$ is the injected 4-momentum in the lab frame $\mathcal{K}$,
  and $q = p'-p-\Omega$.
This expression takes the form $\langle iT_{\rm AA}\rangle = i\mathcal{M}_{\rm AA}\times(2\pi )^{4}\delta^{4}(p'+k'-p-k-\Omega)$,
  where $\mathcal{M}_{\rm AA} = -4m_{\psi_{1}}m_{\psi_{2}}\mathcal{V}_{\rm AA}$ is determined by the Born approximation.
Consequently, the pseudovector/pseudovector interaction potential in the momentum space is 
\begin{equation}
  V_{\rm AA}(q) = g_{\rm A}^{\psi_{1}}g_{\rm A}^{\psi_{2}}
                  \frac{\mathcal{O}_{2}+\frac{M^{2}}{m_{\rm X}^{2}}\mathcal{O}_{3}}{q^{2}-m_{\rm X}^{2}},
  \label{equ15}
\end{equation} 
  and the inverse Lorentz transformation gives the potential in the position space,
\begin{equation}
  V_{\rm AA}(x) \approx -g_{\rm A}^{\psi_{1}}g_{\rm A}^{\psi_{2}}(\mathcal{V}_{2}+\frac{M^{2}}{m_{\rm X}^{2}}\mathcal{V}_{3})
               e^{-r/\mu}e^{-i\omega t}
  \label{equ16}
\end{equation} 
  where $\mu^{-2} = m_{\rm X}^{2}-\omega^{2}\gamma^{2}$.

The distance dependence of the exponential factor when $\omega = 0$, $0<\gamma\omega<m_{\rm X}$, $\gamma\omega = m_{\rm X}$, and $\gamma\omega > m_{\rm X}$
  at $t=0$ is shown in Fig. \ref{FigureS1} (a), (b), (c), and (d), respectively.
When $\omega$ equals zero, Eq. \ref{equ16} degenerates to the static form, i.e., $\mu = \lambda$ and the potential is time-independent.
When $0<\gamma\omega<m_{\rm X}$, the effective force range $\mu$ becomes larger as $\omega$ increases.
In the near-threshold regime, i.e., $\gamma\omega \approx m_{\rm X}$, the exponential factor tends to unity.
While as $\gamma\omega$ exceeds the mass of the new bosons,
  the force range becomes imaginary,
  and the exotic interactions no longer exponentially decay but rather oscillate as the distance $r$ grows.

\subsection{Other types of exotic interactions}
Given that we have proved that the force range of $V_{\rm AA}$ can be enlarged when one of the spins oscillates at a frequency $\omega$,
  here we extend the conclusion to all types of exotic interactions.
Usually three types of new bosons are considered in the study of exotic interactions:
  axions/axionlike particles $\phi$, $Z'$ bosons, and paraphotons $A'$.
In the quantum field theory (QFT) the interaction Hamiltonian contributed by them are 
\begin{subequations}
  \begin{equation}
    \mathcal{H}_{\rm I}^{\phi} = \phi\bar{\psi}(ig_{\rm p}^{\psi}\gamma^{5}+g_{\rm s}^{\psi})\psi,
    \label{equ4a}
  \end{equation}
  \begin{equation}
    \mathcal{H}_{\rm I}^{Z'} = Z'_{\mu}\bar{\psi}\gamma^{\mu}(g_{\rm A}^{\psi}\gamma^{5}+g_{\rm V}^{\psi})\psi,
    \label{equ4b}
  \end{equation}
  \begin{equation}
    \mathcal{H}_{\rm I}^{A'} = \frac{v_{h}}{\Lambda^{2}}P_{\mu\nu}\bar{\psi}\sigma^{\mu\nu}[i\textrm{Im}(C_{\psi})\gamma^{5}+\textrm{Re}(C_{\psi})]\psi,
    \label{equ4c}
  \end{equation}
\end{subequations}
  respectively,
  where $\psi$ denotes the fermionic field, 
  $P_{\mu\nu} = \partial_{\mu}A'_{\nu} - \partial_{\nu}A'_{\mu}$ is the field-strength tensor of the paraphoton,
  $g_{\rm p}^{\psi}$, $g_{\rm s}^{\psi}$, $g_{\rm A}^{\psi}$, $g_{\rm V}^{\psi}$, $\textrm{Im}(C_{\psi})$, and $\textrm{Re}(C_{\psi})$ 
  parameterize the pseudoscalar, scalar, pseudovector, vector, pseudotensor, and tensor interaction strengths, respectively \cite{1984Moody,2019Fadeev,2025Cong}.
By applying Eq. 3 in the main text to all types of interactions, we can get the interaction elements of the S-matrix,
\begin{subequations}
  \begin{equation}
    \begin{aligned}
      \langle iT_{\rm pp}\rangle = &iM^{2}g_{\rm p}^{\psi_{1}}g_{\rm p}^{\psi_{2}}
                                  \frac{\mathcal{O}_{3}}{q^{2}-m_{\rm X}^{2}}
                                  \times(2\pi )^{4}\delta^{4}(p'+k'-p-k-\Omega),
      \label{equ8}
    \end{aligned}
  \end{equation}
  \begin{equation}
    \begin{aligned}
      \langle iT_{\rm ps}\rangle = &-2iMm_{\psi_{2}}g_{\rm p}^{\psi_{1}}g_{\rm s}^{\psi_{2}}
                                  \frac{\mathcal{O}_{9}+\mathcal{O}_{10}}{q^{2}-m_{\rm X}^{2}}
                                  \times(2\pi )^{4}\delta^{4}(p'+k'-p-k-\Omega),
      \label{equ8}
    \end{aligned}
  \end{equation}
  \begin{equation}
    \begin{aligned}
      \langle iT_{\rm ss}\rangle = &-4im_{\psi_{1}}m_{\psi_{2}}g_{\rm p}^{\psi_{1}}g_{\rm s}^{\psi_{2}}
                                  \frac{\mathcal{O}_{1}}{q^{2}-m_{\rm X}^{2}}
                                  \times(2\pi )^{4}\delta^{4}(p'+k'-p-k-\Omega),
      \label{equ8}
    \end{aligned}
  \end{equation}
  \begin{equation}
    \begin{aligned}
      \langle iT_{\rm AA}\rangle = &-ig_{\rm A}^{\psi_{1}}g_{\rm A}^{\psi_{2}}4m_{\psi_{1}}m_{\psi_{2}}
                                  \frac{\mathcal{O}_{2}+\frac{M^{2}}{m_{\rm X}^{2}}\mathcal{O}_{3}}{q^{2}-m_{\rm X}^{2}}
                                  \times(2\pi )^{4}\delta^{4}(p'+k'-p-k-\Omega),
      \label{equ8}
    \end{aligned}
  \end{equation}
  \begin{equation}
    \begin{aligned}
      \langle iT_{\rm AV}\rangle = &ig_{\rm A}^{\psi_{1}}g_{\rm V}^{\psi_{2}}4m_{\psi_{1}}m_{\psi_{2}}
                                  \frac{\frac{M}{m_{\psi_{2}}}\mathcal{O}_{11}+\mathcal{O}_{12}+\mathcal{O}_{13}}{q^{2}-m_{\rm X}^{2}}
                                  \times(2\pi )^{4}\delta^{4}(p'+k'-p-k-\Omega),
      \label{equ8}
    \end{aligned}
  \end{equation}
  \begin{equation}
    \begin{aligned}
      \langle iT_{\rm VV}\rangle =  &ig_{\rm V}^{\psi_{1}}g_{\rm V}^{\psi_{2}}4m_{\psi_{1}}m_{\psi_{2}}
                                    \frac{\mathcal{O}_{1}}{q_{\rm s}^{2}-m_{\rm X}^{2}}
                                    \times(2\pi )^{4}\delta^{4}(p'+k'-p-k)\\
                                   -&ig_{\rm V}^{\psi_{1}}g_{\rm V}^{\psi_{2}}
                                    \frac{|\bm{q}|^{2}\mathcal{O}_{2}-M^{2}\mathcal{O}_{3}}{q^{2}-m_{\rm X}^{2}}
                                    \times(2\pi )^{4}\delta^{4}(p'+k'-p-k-\Omega),
      \label{equ8}
    \end{aligned}
  \end{equation}
  \begin{equation}
    \begin{aligned}
      \langle iT_{\rm TT}\rangle =  &\frac{4i\textrm{Re}(C_{\psi 1})\textrm{Re}(C_{\psi 2})v_{h}^{2}}{\Lambda^{4}}
                                    \frac{|\bm{q_{s}}|^{4}\mathcal{O}_{1}}{q_{\rm s}^{2}-m_{\rm X}^{2}}
                                    \times(2\pi )^{4}\delta^{4}(p'+k'-p-k)\\
                                   -&\frac{16i\textrm{Re}(C_{\psi 1})\textrm{Re}(C_{\psi 2})v_{h}^{2}}{\Lambda^{4}}m_{\psi_{1}}m_{\psi_{2}}
                                    \frac{|\bm{q}|^{2}\mathcal{O}_{2}-M^{2}\mathcal{O}_{3}}{q^{2}-m_{\rm X}^{2}}
                                    \times(2\pi )^{4}\delta^{4}(p'+k'-p-k-\Omega),
      \label{equ8}
    \end{aligned}
  \end{equation}
  \begin{equation}
    \begin{aligned}
      \langle iT_{\rm \widetilde{T}T}\rangle =  
                                    &\frac{8i\textrm{Im}(C_{\psi 1})\textrm{Re}(C_{\psi 2})v_{h}^{2}}{\Lambda^{4}}
                                    \times\frac{m_{\psi_{1}}|\bm{q}|^{2}(\mathcal{O}_{9}+\mathcal{O}_{10}) + 2M^{3}\mathcal{O}_{14p+15}}{q^{2}-m_{\rm X}^{2}}
                                    \times(2\pi )^{4}\delta^{4}(p'+k'-p-k-\Omega)\\
      \label{equ8}
    \end{aligned}
  \end{equation}
  \begin{equation}
    \begin{aligned}
      \langle iT_{\rm \widetilde{T}\widetilde{T}}\rangle =  
                                    &\frac{16i\textrm{Im}(C_{\psi 1})\textrm{Im}(C_{\psi 2})v_{h}^{2}}{\Lambda^{4}}
                                    m_{\psi_{1}}m_{\psi_{2}}M^{2}
                                    \frac{\mathcal{O}_{3}}{q^{2}-m_{\rm X}^{2}}
                                    \times(2\pi )^{4}\delta^{4}(p'+k'-p-k-\Omega),
      \label{equ8}
    \end{aligned}
  \end{equation}
\end{subequations} 
  where $q = p'-p-\Omega$ and $q_{\rm s} = p'-p$.
By comparing the S-matrix element to the Born approximation formula from nonrelativistic quantum mechanics,
  we can determine the potential $V(x)$ generated by the new bosons:
\begin{subequations}
  \begin{equation}
    V_{\rm pp} = g_{\rm p}^{1}g_{\rm p}^{2}\frac{M^{2}}{4m_{\psi_{1}}m_{\psi_{2}}}\mathcal{V}_{3}e^{-r/\mu}e^{-i\omega t},
  \end{equation}
  \begin{equation}
    V_{\rm ps} = g_{\rm p}^{1}g_{\rm s}^{2}\frac{M}{2m_{\psi_{1}}}(\mathcal{V}_{9}+\mathcal{V}_{10})e^{-r/\mu}e^{-i\omega t},
  \end{equation}
  \begin{equation}
    V_{\rm ss} = -g_{\rm s}^{1}g_{\rm s}^{2}\mathcal{V}_{1}e^{-r/\lambda},
  \end{equation}
  \begin{equation}
    V_{\rm AA} = -g_{\rm A}^{1}g_{\rm A}^{2}(\mathcal{V}_{2}+\frac{M^{2}}{m_{\rm X}^{2}}\mathcal{V}_{3})e^{-r/\mu}e^{-i\omega t},
  \end{equation}
  \begin{equation}
    V_{\rm AV} = g_{\rm A}^{1}g_{\rm V}^{2}(\frac{M}{2m_{2}}\mathcal{V}_{11}+\mathcal{V}_{12}+\mathcal{V}_{13})e^{-r/\mu}e^{-i\omega t},
  \end{equation}
  \begin{equation}
    \begin{aligned}
      V_{\rm VV} &= g_{\rm V}^{1}g_{\rm V}^{2}\mathcal{V}_{1}e^{-r/\lambda}
                    +g_{\rm V}^{1}g_{\rm V}^{2}\frac{\mu^{-2}\mathcal{V}_{2}+M^{2}\mathcal{V}_{3}}{4m_{\psi_{1}}m_{\psi_{2}}}e^{-r/\mu}e^{-i\omega t},
    \end{aligned}
  \end{equation}
  \begin{equation}
    \begin{aligned}
      V_{\rm TT} &= \frac{\textrm{Re}(C_{\psi 1})\textrm{Re}(C_{\psi 2})v_{h}^{2}}{\Lambda^{4}}
                  \times\frac{m_{\rm X}^{4}}{m_{\psi_{1}}m_{\psi_{2}}}\mathcal{V}_{1}e^{-r/\lambda}
                  + \frac{4\textrm{Re}(C_{\psi 1})\textrm{Re}(C_{\psi 2})v_{h}^{2}}{\Lambda^{4}}
                  (\mu^{-2}\mathcal{V}_{2}+M^{2}\mathcal{V}_{3})e^{-r/\mu}e^{-i\omega t},\\
    \end{aligned}
  \end{equation}
  \begin{equation}
    \begin{aligned}
      V_{\rm \widetilde{T}T} &= \frac{4\textrm{Im}(C_{\psi 1})\textrm{Re}(C_{\psi 2})v_{h}^{2}}{\Lambda^{4}}
             [\frac{M\mu^{-2}}{2m_{\psi_{2}}}(\mathcal{V}_{9}+\mathcal{V}_{10})+M^{2}\mathcal{V}_{14p+15}]e^{-r/\mu}e^{-i\omega t},
    \end{aligned}
  \end{equation}
  \begin{equation}
    \begin{aligned}
      V_{\rm \widetilde{T}\widetilde{T}} = 4\textrm{Im}(C_{\psi 1})\textrm{Im}(C_{\psi 2})\frac{v_{h}^{2}}{\Lambda^{4}}
                                           M^{2}\mathcal{V}_{3}e^{-r/\mu}e^{-i\omega t},
    \end{aligned}
  \end{equation}
\end{subequations}
  where $\mu^{-2} = m_{\rm X}^{2}-\omega^{2}\gamma^{2}$.
The force ranges of the spin-dependent components in all types of exotic interactions are enlarged.
However, for $V_{\rm ss}$, $V_{\rm VV}$, and $V_{\rm TT}$, the spin-independent components, i.e., the $\mathcal{V}_{1}$ terms, remain unchanged.
\end{widetext}

\section{Calculation of the pseudomagnetic field}

In this section we provide the details of the calculation of the pseudomagnetic field in the near-threshold regime.
For a given precession frequency $\omega$ and Lorentz factor $\gamma$
  the average pseudomagnetic field generated by the spin source spins on the spin sensor spin is
\begin{equation}
  \bm{B}_{\rm exo}(\omega,\gamma) =  \frac{1}{V_{2}}\int_{V_{2}} d\bm{r}_{2} \int_{V_{1}} d\bm{r}_{1} \rho_{1}\bm{b}_{\rm exo}(\omega,\gamma),
\end{equation}
  where $\bm{r}_{1}$ is the position of the source spin, $\bm{r}_{2}$ is the position of the sensor spin, $V_{1}$ is the volume of the spin source, $V_{2}$ is the spin sensor, 
  $\rho_{1}$ is the spin density of the spin source, and $\bm{b}_{\rm exo}$ is the pseudomagnetic field between a source spin and a sensor spin.
However, in practice, the precession frequency $\omega$ of the source spins is not pure-tone, but has a distribution.
There are two reasons:
(1) due to the Doppler effect, the precession frequency $\omega(\bm{\beta},\omega_{0})$ is $\omega_{0}\eta_{\rm D}$,
  where $\eta_{\rm D} = (1-\bm{\widehat{\beta}\cdot\widehat{r}})/(1+\bm{\widehat{\beta}\cdot\widehat{r}})$ is the Doppler coefficient 
  and $\bm{\beta}$ is the velocity of the thermal motion of the atoms with $\langle|\bm{\beta}|\rangle = \sqrt{3k_{\rm B}T/2m_{\rm Fe}}$;
(2) the phase noises of microwave sources makes $\omega_{0}$ itself not pure-tone.
Here we define the quality factor of a microwave source as $Q = \delta\omega_{0}/\omega_{0}$, 
  where $\delta\omega_{0}$ is the linewidth of the output signal of the microwave source.
Consequently, the expression of the pseudomagnetic field should be
\begin{equation}
  \begin{aligned}
    \bm{B}_{\rm exo} = &\int d\omega_{0}S(\omega_{0}) \int d\bm{\beta}p_{1}(\bm{\beta})\\
      &\frac{1}{V_{2}}\int_{V_{2}} d\bm{r}_{2} \int_{V_{1}} d\bm{r}_{1} \rho_{1}\bm{b}_{\rm exo}(\omega(\bm{\beta},\omega_{0}),\gamma(\bm{\beta})),
      \label{Eq12}
  \end{aligned}
\end{equation}
  where $S(\omega_{0})$ is the normalized spectral density of the output signal of the microwave source,
  and $p_{1}(\bm{\beta})$ is the Maxwell-Boltzmann distribution.

\begin{figure}
  \centering
  \includegraphics[width=1\columnwidth]{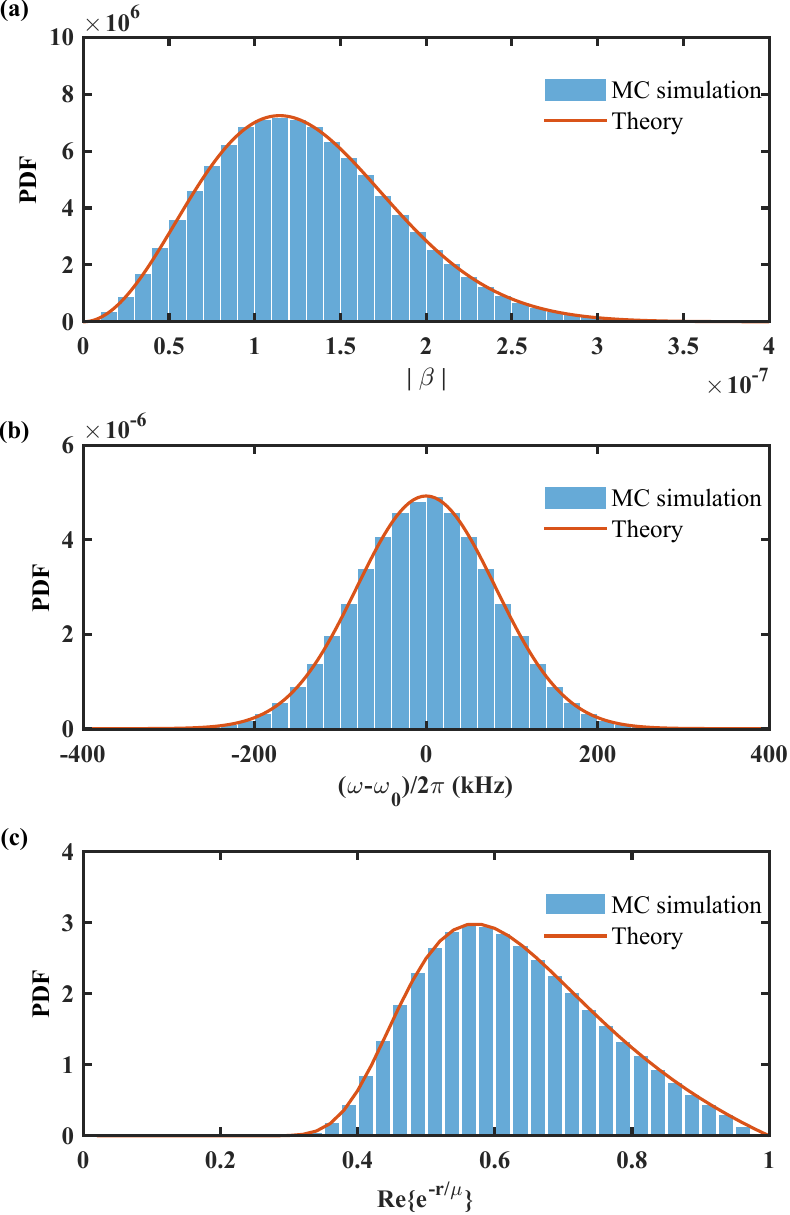}
  \caption{ 
      The probability density function (PDF) of 
      (a) the velocity of the iron atoms in YIG;
      (b) the precession frequency of the electron spins;
      (c) the real part of the exponential factor $e^{-r/\mu}$.
      The orange lines present the theoretical PDF while the blue bars present the result of Monte Carlo simulations.
  }
  \label{FigureS2}
\end{figure}

At 4 K, the velocity of the iron atoms in YIG is around  $10^{-7}$,
  and the distribution of $\omega$ is dominated by the Doppler effect.
The distribution of $|\bm{\beta}|$, $\omega$, and $e^{-r\sqrt{m_{\rm X}^{2}-\gamma^{2}\omega^{2}}}$ 
  with $\omega_{0} = 2\pi\times 500$ GHz, $\lambda = 0.1$ mm, and $r = 0.1$ m 
  is shown in Fig. \ref{FigureS1}(a), (b), and (c), respectively.
The orange lines show the theoretical probability density functions,

\begin{subequations}
  \begin{equation}
    p_{1}(\beta) = \frac{4\pi\beta^{2}}{(\sqrt{\pi}\beta_{0})^{3}}e^{-\beta/\beta_{0}},
  \end{equation}
  \begin{equation}
    \begin{aligned}
      p_{2}(\omega) = \frac{2\omega_{0}}{\sqrt{\pi}\beta_{0}(\omega_{0}+\omega)^{2}}e^{-[\frac{\omega_{0}-\omega}{\beta_{0}(\omega_{0}+\omega)}]^{2}},
    \end{aligned}
  \end{equation}
  \begin{equation}
    \begin{aligned}
      p_{3}(\textrm{Re}\{e^{-r/\mu}\}) 
        = \frac{2\omega_{0}\textrm{ln}x/\omega r^{2}x}{\sqrt{\pi}\beta_{0}(\omega_{0}+\omega)^{2}}e^{-[\frac{\omega_{0}-\omega}{\beta_{0}(\omega_{0}+\omega)}]^{2}},
    \end{aligned}
  \end{equation}
\end{subequations}
  where $\beta_{0} = \sqrt{2k_{\rm B}T/m_{\rm Fe}}$, $x=\textrm{Re}\{e^{-r/\mu}\}$ and $\omega = \sqrt{\omega_{0}^{2}-(\textrm{ln}x/r)^{2}}$.
These results are verified through Monte Carlo (MC) simulations, which are shown as the blue bars in Fig. \ref{FigureS2}

For numerical calculation, Eq. \ref{Eq12} should be discretized into the following form:
\begin{equation}
  \begin{aligned}
    \bm{B}_{\rm exo} = &2\pi\sum_{l}^{N_{1}} \Delta\omega_0 S(\omega_{0}^{l})\\
      &\sum_{m}^{N_{2}} \Delta\theta \sin\theta^{m} \sum_{n}^{N_{3}} \Delta\beta (n\beta)^{2} p_{1}(\bm{\beta}^{mn})\\
      &\frac{\rho_{1}V_{1}}{M_{1}M_{2}}\sum_{i}^{M_{1}}\sum_{j}^{M_{2}}
      \bm{b}_{\rm exo}^{ij}(\omega(\bm{\beta}^mn,\omega_{0}^{l}),\gamma(\bm{\beta}^{mn})),
  \end{aligned}
\end{equation}
  where $\Delta\omega_{0}$, $\Delta\theta$, and $\Delta\beta$ represent the integration steps of $\omega_{0}$, $\theta$, and $\beta$, respectively,
  $\omega_{0}^{l} = l\Delta\omega_{0}$, $\theta^{m} = m\Delta\theta$, and $\bm{\beta}^{mn} = n\Delta\beta(0,\sin\theta,\cos\theta)$.
The Monte Carlo method was utilized to calculate the integration over $V_{1}$ and $V_{2}$ 
  since there will be around $10^{24}$ spins in the proposed experimental setup and directly summing them up is impossible.
Therefore, we instead generated $M_{1}$ random points in the spin source and $M_{2}$ random points in the spin sensor,
  calculated the pseudomagnetic field between each point pair,
  and estimated the total pseudomagnetic field from their average value.
Since the integration is not sensitive to the summation upper bounds $N_{1,2,3}$ and $M_{1,2}$, 
  the summation upper bounds were chosen to be as large as the calculation result converges to a certain value.


\begin{thebibliography}{10}
\expandafter\ifx\csname url\endcsname\relax
  \def\url#1{\texttt{#1}}\fi
\expandafter\ifx\csname urlprefix\endcsname\relax\def\urlprefix{URL }\fi
\providecommand{\bibinfo}[2]{#2}
\providecommand{\eprint}[2][]{\url{#2}}

\bibitem{1906Poincare}
\bibinfo{author}{H. Poincare},
\bibinfo{title}{The Milky Way and the theory of gases},
{Pop. Astron. \textbf{14}, 475 (1906)}.
\bibitem{1933Zwicky}
\bibinfo{author}{F. Zwicky},
\bibinfo{title}{The redshift of extragalactic nebulae},
\href{https://doi.org/10.1007/s10714-008-0707-4}
{Helv. Phys. Acta. \textbf{6}, 110 (1933)}.
\bibitem{2023Han}
\bibinfo{author}{C. Han},
\bibinfo{title}{QCD axion dark matter and the cosmic dipole problem}.
\href{https://doi.org/10.1103/PhysRevD.108.015026}
{Phys. Rev. D \textbf{108}, 015026 (2023)}.
\bibitem{2021Abi}
\bibinfo{author}{B. Abi \textit{et al.}},
\bibinfo{title}{Measurement of the positive muon anomalous magnetic moment to 0.46 ppm},
\href{https://doi.org/10.1103/PhysRevLett.126.141801}
{Phys. Rev. Lett. \textbf{126}, 141801 (2021)}.
\bibitem{2021Cazzaniga}
\bibinfo{author}{C. Cazzaniga \textit{et al.}},
\bibinfo{title}{Probing the explanation of the muon (g-2) anomaly and thermal light dark matter with the semi-visible dark photon channel},
\href{https://doi.org/10.1140/epjc/s10052-021-09705-5}
{Eur. Phys. J. C \textbf{81}, 959 (2021)}.
\bibitem{2022Aaltonen}
\bibinfo{author}{T. Aaltonen},
\bibinfo{title}{High-precision measurement of the W boson mass with the CDF II Detector},
\href{https://www.science.org/doi/10.1126/science.abk1781}
{Science \textbf{376}, 170 (2022)}.
\bibitem{2022Thomas}
\bibinfo{author}{A. W. Thomas and X. G. Wang},
\bibinfo{title}{Constraints on the dark photon from parity violation and the W mass},
\href{https://doi.org/10.1103/PhysRevD.106.056017}
{Phys. Rev. D \textbf{106}, 056017 (2022)}.
\bibitem{1984Moody}
\bibinfo{author}{J. E. Moody and F. Wilczek},
\bibinfo{title}{New macroscopic forces?}
\href{https://doi.org/10.1103/PhysRevD.30.130}
{Phys. Rev. D \textbf{30}, 130 (1984)}.
\bibitem{2006Dobrescu}
\bibinfo{author}{B. A. Dobrescu and I. Mocioiu},
\bibinfo{title}{Spin-dependent macroscopic forces from new particle exchange},
\href{https://doi.org/10.1088/1126-6708/2006/11/005}
{J. High Energy Phys. \textbf{11}, 005 (2006)}.
\bibitem{spp}
\bibinfo{title}{See Supplemental Material for additional information about the background on exotic interactions, details of the derivation of the near-threshold effect, and numerical methods, which includes Refs. \cite{1984Moody,2006Dobrescu,2019Fadeev,2025Cong}.}
\bibitem{2019Fadeev}
\bibinfo{author}{P. Fadeev, Y. V. Stadnik, F. Ficek, M. G. Kozlov, V. V. Flambaum, and D. Budker},
\bibinfo{title}{Revisiting spin-dependent forces mediated by new bosons: Potentials in the coordinate-space representation for macroscopic- and atomic-scale experiments},
\href{https://doi.org/10.1103/PhysRevA.99.022113}
{Phys. Rev. A \textbf{99}, 022113 (2019)}.
\bibitem{2025Cong}
\bibinfo{author}{L. Cong \textit{et al.}},
\bibinfo{title}{Spin-dependent exotic interactions},
\href{https://doi.org/10.1103/RevModPhys.97.025005}
{Rev. Mod. Phys. \textbf{97}, 025005 (2025)}.
\bibitem{2013Tullney}
\bibinfo{author}{K. Tullney \textit{et al.}},
\bibinfo{title}{Constraints on Spin-Dependent Short-Range Interaction between Nucleons},
\href{http://dx.doi.org/10.1103/PhysRevLett.111.100801}
{Phys. Rev. Lett. \textbf{111}, 100801 (2013)}.
\bibitem{2019Kim}
\bibinfo{author}{Y. J. Kim, P-H Chu, I. Savukov, and S. Newman },
\bibinfo{title}{Experimental limit on an exotic parity-odd spin- and velocity-dependent interaction using an optically polarized vapor},
\href{https://doi.org/10.1038/s41467-019-10169-1}
{Nat. Commun. \textbf{11}, 2245 (2019)}.
\bibitem{2018Rong}
\bibinfo{author}{X. Rong \textit{et al.}},
\bibinfo{title}{Searching for an exotic spin-dependent interaction with a single electron-spin quantum sensor},
\href{https://doi.org/10.1038/s41467-018-03152-9}
{Nat. Commun. \textbf{9}, 739 (2018)}.
\bibitem{2021Jiao}
\bibinfo{author}{M. Jiao, M. Guo, X. Rong, Y.-F. Cai, and J. Du},
\bibinfo{title}{Experimental Constraint on an Exotic Parity-Odd Spin- and Velocity-Dependent Interaction with a Single Electron Spin Quantum Sensor},
\href{https://doi.org/10.1103/PhysRevLett.127.010501}
{Phys. Rev. Lett. \textbf{127}, 010501 (2021)}.
\bibitem{2011Hoedl}
\bibinfo{author}{S. A. Hoedl, S. M. Fleischer, E. G. Adelberger, and B. R. Heckel},
\bibinfo{title}{Improved Constraints on an Axion-Mediated Force},
\href{http://dx.doi.org/10.1103/PhysRevLett.106.041801}
{Phys. Rev. Lett. \textbf{106}, 041801 (2011)}.
\bibitem{2015Terrano}
\bibinfo{author}{W. A. Terrano, E. G. Adelberger, J. G. Lee, and B. R. Heckel},
\bibinfo{title}{Short-Range, Spin-Dependent Interactions of Electrons: A Probe for Exotic Pseudo-Goldstone Bosons},
\href{http://dx.doi.org/10.1103/PhysRevLett.115.201801}
{Phys. Rev. Lett. \textbf{115}, 201801 (2015)}.
%
\bibitem{2024Tilburg}
\bibinfo{author}{K. V. Tilburg},
\bibinfo{title}{Wake forces in a background of quadratically coupled mediators},
\href{https://doi.org/10.1103/PhysRevD.109.096036}
{Phys. Rev. D \textbf{109}, 096036 (2024)}.
%
\bibitem{2023Ghosh}
\bibinfo{author}{M. Ghosh, Y. Grossman, W. Tangarife, X.-J. Xu, and B. Yu},
\bibinfo{title}{Neutrino forces in neutrino backgrounds},
\href{https://doi.org/10.1007/JHEP02(2023)092}
{J. High Energ. Phys. \textbf{2023}, 92 (2023)}.
\bibitem{2015Kawasaki}
\bibinfo{author}{M. Kawasaki, K. Saikawa, and T. Sekiguchi},
\bibinfo{title}{Axion dark matter from topological defects},
\href{https://doi.org/10.1103/PhysRevD.91.065014}
{Phys. Rev. D \textbf{91}, 065014 (2015)}.
\bibitem{2016Borsanyi}
\bibinfo{author}{S. Borsanyi \textit{et al.}},
\bibinfo{title}{Calculation of the axion mass based on high-temperature lattice quantum chromodynamics},
\href{https://doi.org/10.1038/nature20115}
{Nature \textbf{539}, 69–71 (2016)}.
\bibitem{2020Crescini}
\bibinfo{author}{N. Crescini \textit{et al.}},
\bibinfo{title}{Axion Search with a Quantum-Limited Ferromagnetic Haloscope},
\href{https://doi.org/10.1103/PhysRevLett.124.171801}
{Phys. Rev. Lett. \textbf{124}, 171801 (2020)}.
\bibitem{1982Caves}
\bibinfo{author}{C. M. Caves},
\bibinfo{title}{Quantum limits on noise in linear amplifiers},
\href{https://doi.org/10.1103/PhysRevD.26.1817}
{Phys. Rev. D \textbf{26}, 1817 (1982)}.
\bibitem{2017Ficek}
\bibinfo{author}{F. Ficek, D. F. J. Kimball, M. G. Kozlov, N. Leefer, S. Pustelny, and D. Budker},
\bibinfo{title}{Constraints on exotic spin-dependent interactions between electrons from helium fine-structure spectroscopy},
\href{https://doi.org/10.1103/PhysRevA.95.032505}
{Phys. Rev. A \textbf{95}, 032505 (2017)}.
\bibitem{2022Fadeev}
\bibinfo{author}{P. Fadeev, F. Ficek, M. G. Kozlov, D. Budker, and V. V. Flambaum},
\bibinfo{title}{Pseudovector and pseudoscalar spin-dependent interactions in atoms},
\href{https://doi.org/10.1103/PhysRevA.105.022812}
{Phys. Rev. A \textbf{105}, 022812 (2022)}.

\end{thebibliography}
\end{document}